\documentclass[aps,pra,final,letterpaper,10pt,twocolumn,longbibliography,floats,showpacs,amsmath,amsfonts,amssymb,superscriptaddress]{revtex4-1}

\usepackage{graphicx}
\usepackage{bm}
\usepackage{bbm}
\usepackage{color}
\usepackage{epstopdf}
\usepackage{amsmath}
\usepackage{amssymb}
\usepackage{ulem}
\usepackage[urlcolor=blue,colorlinks=true,citecolor=blue,linkcolor=blue,pdfstartview={FitH},bookmarks=false]{hyperref}
\usepackage{todonotes}
\usepackage{ulem}

\begin{document}
\title{Dynamical quantum phase transitions in a mesoscopic superconducting  system}

\author{K. Wrze\'sniewski}
\author{I. Weymann}
\affiliation{Institute of Spintronics and Quantum Information, Faculty of Physics, A. Mickiewicz University, 61-614 Pozna\'n, Poland}

\author{N. Sedlmayr}
\author{T. Doma\'nski}
\affiliation{Institute of Physics, M. Curie-Sk\l{}odowska University, 20-031 Lublin, Poland}

\date{\today}

\begin{abstract}
We inspect signatures of dynamical quantum phase transitions driven by two types of quenches acting on a correlated quantum dot embedded between superconducting and metallic reservoirs. Under stationary conditions the proximity induced on-dot pairing, combined with strong Coulomb repulsion, prefers the quantum dot to be either in the singly occupied (spinful) or BCS-type (spinless) ground state configuration. We study the time evolution upon traversing such a phase boundary due to quantum quenches by means of the time-dependent numerical renormalization group approach, revealing non-analytic features in the low-energy return rate. Quench protocols can be realized in a controllable manner and we are confident that detection of this {\it dynamical singlet-doublet phase transition} would be feasible by charge tunnelling spectroscopy.
\end{abstract}

\maketitle

{\it Motivation. --}
Understanding the evolution of quantum many-body systems away from equilibrium is currently a topic of intensive investigations in condensed matter and ultracold atom physics, where the dynamical critical phenomena are driven by time-dependent perturbations~\cite{Polkovnikov-2011}. In particular, dynamical quantum phase transitions (DQPTs)~\cite{Heyl-2013} can appear at critical times $t_{c}$ following a quantum quench. Usually, this occurs for quenches across equilibrium phase transition boundaries, although this is neither a necessary nor sufficient condition~\cite{Vajna2014}. DQPTs manifest themselves by non-analytic (cusp-like) features appearing in the return rate function~\cite{Heyl-2018}, in analogy to a static (classical or quantum) phase transitions revealing itself in the free energy at a critical temperature, magnetic field, pressure, etc. DQPTS have also been generalised to non ground state initial conditions with mixed results~\cite{Abeling2016,Bhattacharya2017a,Heyl2017,Mera2017,Sedlmayr2018b,Lang2018a,Lang2018,Hou2020b}, and in driven systems~\cite{Sharma2014,Kosior2018,vanCaspel2019,Yang2019,Arze2020,Hamazaki2020,Jafari2020,Kennes2020,Bandyopadhyay2021a}. So far, the studies of critical properties~\cite{Heyl2015,Zunkovic2018}, dynamical order parameters~\cite{Canovi2014a,Budich2016}, and spontaneously broken symmetries~\cite{Heyl2014,Kosior2018a} have been mostly addressed in bulk systems, often in either spin chains~\cite{Karrasch2013,Andraschko2014,Azimi2016,Halimeh2017,Karrasch2017,Gurarie2019,Halimeh2018} or topological insulators and superconductors~\cite{Vajna2015,Hegde2015,Schmitt2015,Sedlmayr2018,Mishra2018,Sun2018,Hu2019,Mendl2019,Sedlmayr2019a,Maslowski2020}. Some studies have also addressed the superconducting transition~\cite{Nori-2021,Galitski-2021}. On the other hand, empirical evidence for DQPTs has been reported only in a few cases, mainly using trapped ion systems \cite{Jurcevic2017}, although alternative suggestions for measuring DQPTs exist~\cite{Heyl2014,Heyl2015,Huang2016,Heyl2018,Chen2020,Nie2020}.

\begin{figure}[b]
\includegraphics[width=0.8\columnwidth]{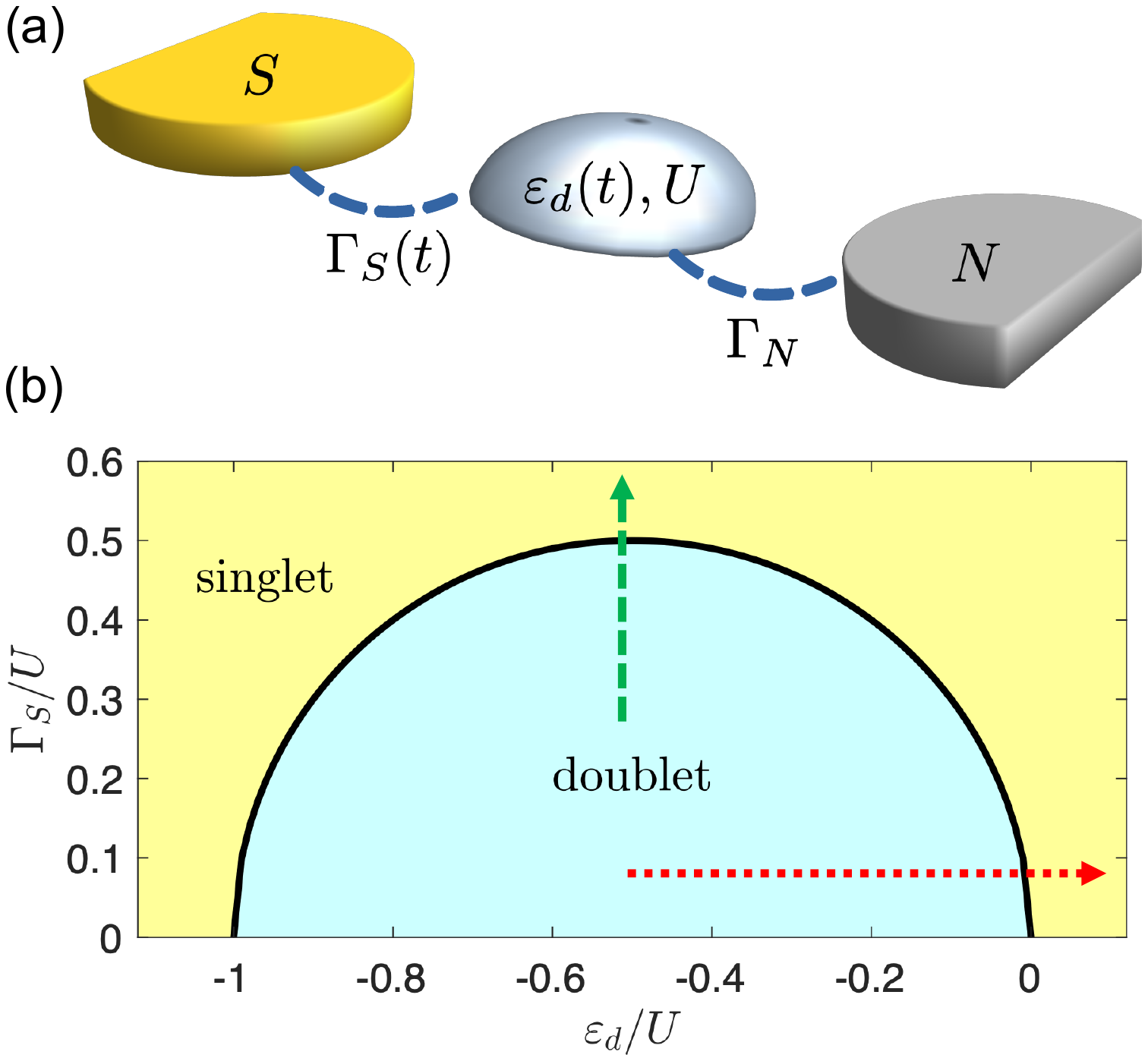}
\caption{(a) Schematic of a correlated quantum dot sandwiched between superconducting (S) and normal (N) leads. A dynamical singlet-doublet transition can originate from a quench imposed either on the coupling strength $\Gamma_{S}(t)$ or the energy level position $\varepsilon_{d}(t)$ when crossing through the phase boundary, which is marked with a solid line in (b). The dashed [dotted] arrow in (b) indicates the direction of the
quench performed in $\Gamma_{S}(t)$ [$\varepsilon_{d}(t)$] studied in this paper.}
\label{scheme}
\end{figure}

While clear-cut signatures of static and dynamical phase transitions are observable solely in the thermodynamic limit, it has been recently shown \cite{Puebla-2020} that DQPTs can also be realized in systems comprising of a limited (finite) number of constituents. Indeed such a result is supported by the appearance of signatures of DQPTs in finite-size analyses~\cite{Heyl_etal-2020}. Here, we propose another realm for a feasible dynamical quantum phase transition in a finite system, which could be realized in hybrid nanostructures. As a specific example, we consider a correlated quantum dot (QD) 
at the interface between superconducting (S) and normal (N) leads, as depicted in Fig.~\ref{scheme}(a). The competition of the superconducting proximity effect with the Coulomb repulsion can induce a changeover in the quantum dot ground state between the singly occupied configuration and a coherent superposition of the empty and doubly occupied (BCS-type) state. We argue that the signatures of the dynamical transition between these qualitatively different configurations would be observable in the subgap charge transport properties of the considered heterostructure. This may open a new route towards a controllable realization of DQPTs in various nanoscale systems, where the device  parameters, i.e.~the quantum dot energy levels and their couplings to external reservoirs, can be  tuned  with unprecedented precision~\cite{Franke-2018,Klinovaja-2021}

{\it General formulation. --}
To outline the underlying idea of a DQPT let us consider a system, initially described by the Hamiltonian $\hat{H}_{0}$, whose ground state obeys the Schr\"odinger equation, $\hat{H}_{0} \left| \Psi_{0} \right> = E_{0} \left| \Psi_{0} \right>$.
At time $t=0$, the Hamiltonian is suddenly changed, $\hat{H}_{0} \rightarrow \hat{H}$, causing the evolution $i \frac{d}{dt}\left| \Psi(t) \right> = \hat{H} \left| \Psi(t) \right>$, which results in the time-dependent state $\left| \Psi(t) \right> = e^{-it\hat{H}} \left| \Psi_{0} \right>$. Fidelity (similarity) of the states $| \Psi_{0} \rangle$ and $|\Psi(t)\rangle$ at an arbitrary time $t\geq 0$ can be characterized by the Loschmidt amplitude, $\left< \Psi_{0} | \Psi(t) \right>  =  \big< \Psi_{0}| 
e^{-it\hat{H}} | \Psi_{0} \big>$.

The squared absolute value of the Loschmidt amplitude, $L(t)=\big| \big< \Psi_{0}| e^{-it\hat{H}} | \Psi_{0} \big>\big|^2$, referred to as the Loschmidt echo, can be regarded as dual to the partition function ${\cal{Z}} 
= \big< e^{-\beta \hat{H}}\big>$ in statistical physics, where the inverse 
temperature  ${\beta = 1/k_{B}T}$ is replaced by the imaginary time $it$. Furthermore, 
the free energy, $F(T)=-k_{B}T\ln{{\cal{Z}}}$, 
is equivalent to the return rate $\lambda(t)$ defined via $L(t) \equiv e^{-N \lambda(t)}$, where $N$ stands for the number of degrees of freedom. The usual critical temperature $T_{c}$, referring to a discontinuity of the free energy or its derivatives  $\lim_{T\rightarrow T_{c}} F(T)$, 
is hence equivalent to the critical time $t_{c}$, at which the non-analyticities occur 
in the return rate $\lim_{t\rightarrow t_{c}} \lambda(t)$. For simple systems a sequence of critical points equally spaced along the time axis occurs, however, non-periodic behavior is also possible~\cite{Heyl-2018,Maslowski2020}, and likely generic.

{\it Microscopic model. --}
Let us  present the microscopic scenario, in which the quantum quench $\hat{H}_{0} \rightarrow \hat{H}$ qualitatively affects the properties of our heterostructure. The correlated quantum dot contacted with the superconducting and normal leads (Fig.~\ref{scheme}) can be described by the Hamiltonian
\begin{eqnarray}
\hat{H} =  \hat{H}_{QD}  + \sum_{\beta} \big( \hat{H}_{\beta} + 
\hat{V}_{\beta - QD} \big) ,
\label{model}
\end{eqnarray}
with $\beta\in\{N,S\}$ referring to the metallic and superconducting reservoirs respectively. 
The correlated quantum dot is described by $\hat{H}_{QD}=\sum_{\sigma} \varepsilon_{d}(t) 
\hat{d}^{\dagger}_{\sigma} \hat{d}_{\sigma} + U\hat{n}_{\uparrow} \hat{n}_{\downarrow}$,
where $\hat{d}_{\sigma}$ ($\hat{d}^{\dagger}_{\sigma}$) stands for the annihilation (creation) operator of spin $\sigma=\uparrow, \downarrow$ electrons whose energy 
$\varepsilon_{d}(t)$ depends on time, and $U>0$ is the Coulomb potential of repulsive 
interactions in the dot. The macroscopic superconductor is assumed to be of BCS type: 
$\hat{H}_{S} \!=\!\sum_{{\bf k},\sigma}  \xi_{S {\bf k}} \hat{c}_{S {\bf k}\sigma}^{\dagger}  
\hat{c}_{S{\bf k}\sigma} \!-\! \sum_{\bf k} \Delta  \big( \hat{c}_{S{\bf k} \uparrow} ^{\dagger}
\hat{c}_{S -{\bf k} \downarrow}^{\dagger} + \hat{c} _{S -{\bf k} \downarrow} \hat{c}_{S{\bf k}
\uparrow }\big)$, with an isotropic pairing gap $\Delta$ and dispersion relation 
$\xi_{S{\bf k}}$. The normal lead will be simply treated as a free fermion gas,
$\hat{H}_{N} \!=\! \sum_{{\bf k},\sigma} \xi_{N {\bf k}} \hat{c}_{N{\bf k} \sigma}^{\dagger}
\hat{c}_{N{\bf k} \sigma}$.
The hybridization of the localized (QD) and itinerant (N,S) electrons is given by, 
$\hat{V}_{\beta-QD} = \sum_{{\bf k},\sigma} \big( V_{\beta {\bf k}} \; \hat{d}_{\sigma}^{\dagger}
\hat{c}_{\beta {\bf k} \sigma} + V^{*}_{\beta {\bf k}} \; \hat{c}_{\beta {\bf k} \sigma}^{\dagger} 
\hat{d}_{\sigma} \big)$, where  $\hat{c}_{\beta {\bf k}\sigma}$
is the corresponding annihilation operator
and $V_{\beta {\bf k}}$ denotes the tunnelling matrix elements.

From here onward we analyze the low energy properties of such a N-QD-S hybrid structure,
restricted to the region $|\omega| \ll \Delta$, safely inside the pairing gap $\Delta$, 
which in conventional superconductors is usually of the order of meV~\cite{Wernsdorfer-2010}. Under such circumstances, we can impose the auxiliary couplings $\Gamma_{\beta}=\pi 
\sum_{\bf k} |V_{\beta{\bf k}}|^2 \;\delta(\omega \!-\! \xi_{\beta{\bf k}})$, 
assuming them to be energy-independent. The mobile Cooper pairs leaking onto the quantum
dot will induce on-dot pairing. By integrating out the fermionic degrees of freedom 
from outside the pairing gap, such a superconducting proximity effect can be modeled by 
\begin{eqnarray}
\hat{H}_{S} + \hat{V}_{S - QD} \approx  \big( \Gamma_{S} \; \hat{d}^{\dagger}_{\uparrow} 
\hat{d}_{\downarrow}^{\dagger} +\mbox{\rm H.c.} \big),
\label{proximity_effect}
\end{eqnarray}
so that $\Gamma_{S}$ effectively plays the role of a pairing potential. The term \eqref{proximity_effect} competes with the repulsive Coulomb interaction, leading to different characteristic signatures. We shall discuss them briefly in the following. The coupling to the normal lead $\Gamma_{N}$  will be assumed to be much smaller than $\Gamma_{S}$, to guarantee a sufficiently long life-time of the in-gap quasiparticles. Such a situation is customarily encountered in scanning tunneling microscope (STM) measurements, where various nanoscopic objects deposited on surfaces of superconductors are probed by a normal or superconducting tip placed at a secure distance from the impurities.

{\it Equilibrium quantum phase transition. --}
Interplay between the superconducting proximity effect and electron correlations
leads to qualitative changes of the QD ground state \cite{Bauer-2007}. True 
eigenstates of the quantum dot are represented either by the singly occupied 
configurations $\left| \sigma \right>$ (which are degenerate in the absence of a magnetic field) or BCS-type superpositions of the empty and doubly occupied 
states $u\left|0\right>-v\left|\uparrow\downarrow\right>$, 
$v\left|0\right>+u\left|\uparrow\downarrow\right>$. Upon changing the ratio of $\Gamma_{S}/U$ (or the QD level $\varepsilon_{d}$), the ground state can evolve 
between these spinful and spinless configurations. Such a parity changeover is 
manifested by the energy crossing of the in-gap quasiparticles, known as Andreev or Yu-Shiba-Rusinov bound states~\cite{Yu.1965,Shiba.1968,Rusinov.1974}. In the superconducting atomic limit
(when the pairing gap $\Delta$ is the largest energy scale) the quantum phase transition between the doublet and the BCS-type singlet occurs at  
\begin{eqnarray}
\left( \varepsilon_{d} + \frac{U}{2} \right)^{\!2} \!+ \Gamma_{S}^{2} 
= \left( \frac{U}{2}\right)^{\!2}.
\end{eqnarray}
The corresponding phase diagram is presented in Fig.~\ref{scheme}(b). In practical realizations of the superconducting N-QD-S and/or S-QD-S 
heterostructures the impurity levels can be varied by a gate potential, 
changing their even/odd occupancy, which in turn affects the subgap properties \cite{Wernsdorfer-2010}. For instance, in S-QD-S circuits such a
singlet-doublet quantum phase transition is responsible for a reversal of 
the d.c.~Josephson current, the so-called ``$0-\pi$ transition''. As far as  
N-QD-S junctions are concerned, the singlet to doublet crossover becomes smooth~\cite{Jellinggaard-2016} because the subgap quasiparticles acquire a finite life-time due to the coupling $\Gamma_{N}$ to the continuous spectrum of the normal lead electrons. Nonetheless, such a crossover has been
clearly observed experimentally by various groups (see e.g. Ref.~\cite{Nygard-2020} and other references cited therein). The singlet-doublet
transition has also further implications for the magnetic susceptibility and 
pairing field $\big< \hat{d}_{\downarrow} \hat{d}_{\uparrow}\big>$,
and has been predicted to enhance the subgap Kondo effect upon approaching the phase boundary from the doublet side \cite{Zitko-2015,Domanski-2016}.

{\it Quench protocol for the singlet-doublet transition. --} We now propose two possible 
scenarios of the quantum quench $\hat{H}_{0} \longrightarrow \hat{H}$, in which an abrupt 
change of our setup would enforce a transition between the Hamiltonians with singlet and doublet QD ground state configurations.
In the first approach, we impose a variation of the coupling to the superconductor
\begin{eqnarray}
\Gamma_{S}(t) =
\left\{ \begin{array}{ll}
\Gamma_{S0} & \hspace{0.5cm} \mbox{\rm for } t \leq 0  , \\
\Gamma_{S}  & \hspace{0.5cm} \mbox{\rm for } t > 0 ,
\end{array} \right.
\label{abrupt_coupling}
\end{eqnarray}
assuming that the initial ($\Gamma_{S0}$) and final ($\Gamma_{S}$) values are on the opposite sides of the phase transition boundary [indicated with a dashed line in Fig.~\ref{scheme}(b)]. The second type of quantum quench affects the energy level of the QD [marked with a dotted line in Fig.~\ref{scheme}(b)]
\begin{eqnarray}
\varepsilon_{d}(t) =
\left\{ \begin{array}{ll}
\varepsilon_{d} & \hspace{0.5cm} \mbox{\rm for } t \leq 0  , \\
\varepsilon_{d} + V_{G}  & \hspace{0.5cm} \mbox{\rm for } t > 0 .
\end{array} \right.
\label{abrupt_gate}
\end{eqnarray}
In both cases (\ref{abrupt_coupling},\ref{abrupt_gate}) we choose such parameters which guarantee the quantum dot to be initially in the doublet and finally in the singlet configurations. Evolution between these different states (via a sequence of critical times) in a long-time limit terminates in the steady state, owing to relaxation processes in the continuous spectrum of the normal electrode \cite{Wrzesniewski-2021}. Additional sources of such relaxation phenomena could be provided by electronic states existing outside the superconducting gap~\cite{LevyYeyati-2021}, but we neglect their influence here.

\begin{figure}
\includegraphics[width=0.99\columnwidth]{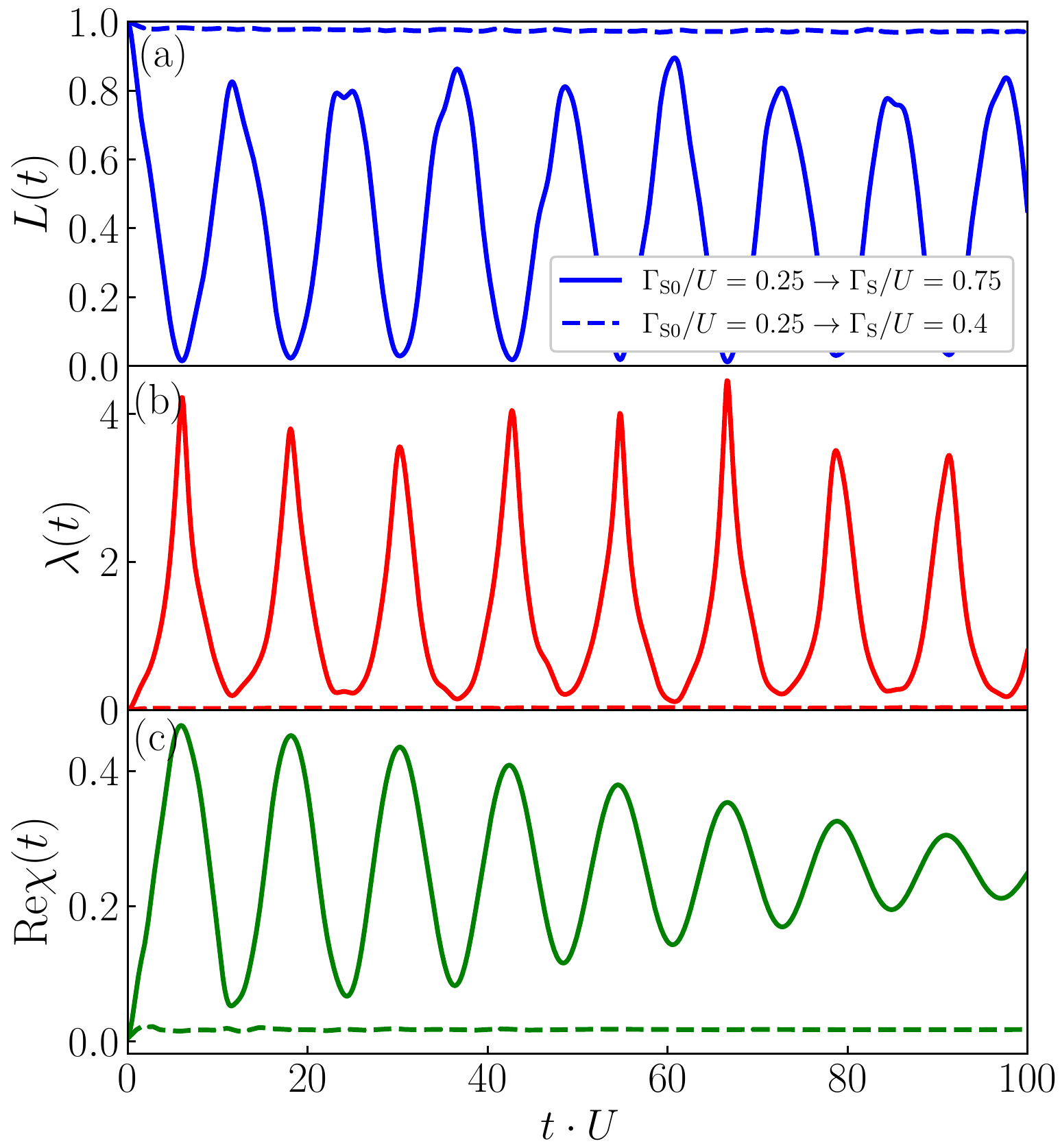}
\caption{(a) The time evolution of the Loschmidt echo, (b) the return rate and (c) the on-dot paring after the quench in the coupling to superconductor $\Gamma_{S}$. The solid lines correspond to the quench across the phase boundary between the doublet and singlet states, as marked with the dashed arrow in Fig.~\ref{scheme}(b). The dashed lines present the results for quench within the doublet phase. The other parameters are: $U=0.1$,
$\varepsilon_d=-0.05$ and $\Gamma_N=0.001$ in units of half the bandwidth.} 
\label{quench_in_Gamma_S}
\end{figure}

\begin{figure}
\includegraphics[width=0.9\columnwidth]{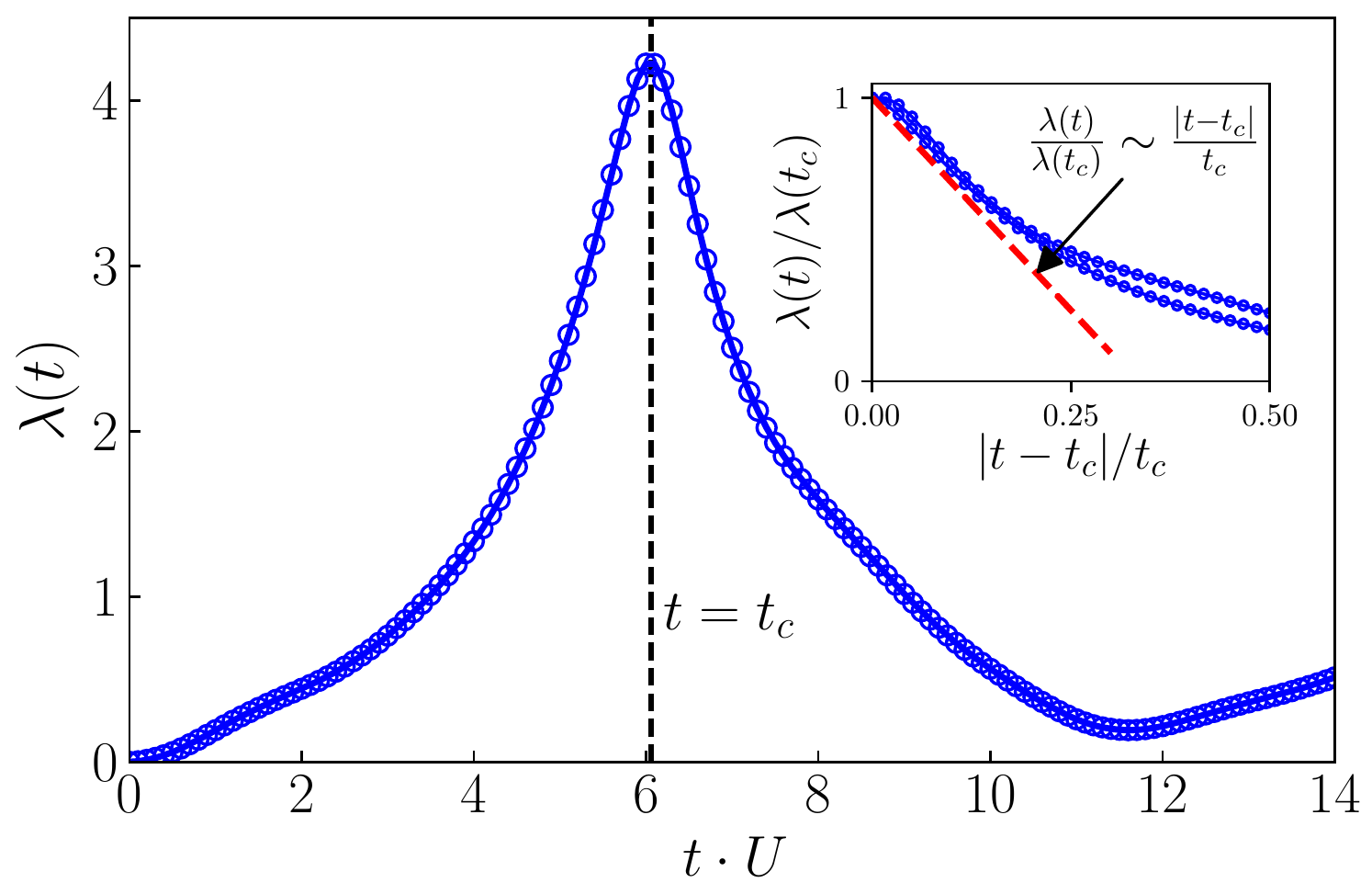}
\caption{The time dependence of the return rate around the first critical time $t_c$
visible in Fig.~\ref{quench_in_Gamma_S}.
The inset presents the dependence of $\lambda(t)/\lambda(t_c)$
on the normalized time $|t-t_c|/t_c$.
The parameters are the same as in Fig.~\ref{quench_in_Gamma_S}.
} 
\label{RR_tc}
\end{figure}

In STM spectroscopy the coupling of an impurity to  superconducting substrate could be varied by manipulating the tip-impurity distance~\cite{Franke-2018}. This allows us to traverse the quantum phase transition by changing the effective exchange interaction \cite{Klinovaja-2021}. As regards the ballistic charge tunneling via metal - quantum dot - superconductor hybrid nanostructures, there is a fairly large flexibility in a controllable variation of both the coupling $\Gamma_{S}$ and the quantum dot energy level $\varepsilon_{d}$. Evidence for the resulting parity (singlet-doublet) crossings has been reported by several groups, for instance using InAs nanowires placed between superconducting Al and metallic Au electrodes~\cite{Nygard-2020,DeFranceschi-2017} and in a carbon nanotube contacted to superconducting Nb and weakly coupled to a normal metal~\cite{Baumgartner-2014}.

{\it Dynamical quantum phase transitions. --}
To accurately describe the dynamical behavior of the system we resort to the numerical renormalization group method (NRG)~\cite{Wilson1975,Bulla2008,NRG_code}. This method has been successfully used to analyze the stationary singlet-doublet transition \cite{Bauer-2007}. Here, we make use of its time-dependent extension \cite{Anders2005,Costi2014,WrzesniewskiWeymann-2019} to address the problem of dynamical quantum phase transitions. The core of NRG is a logarithmic discretization (with parameter $\Lambda$) of the conduction band, which allows one to map the Hamiltonian to a chain-like form. Such a model can be diagonalized in an iterative fashion by keeping an appropriate number of low-energy states $N_K$.
This allows us to find complete many-body eigenbases of the Hamiltonians $\hat{H}_0$ and $\hat{H}$,
$\sum_{nse}|nse\rangle^{\!D}_{0} \,{}^{D}_{\,0}\!\langle nse| \!=\! \mathbbm{\hat{1}}$
and
$\sum_{nse}|nse\rangle^{\!D} \,{}^D \!\langle nse| \!=\! \mathbbm{\hat{1}}$,
respectively, where $s$ denotes an eigenstate at iteration $n$ belonging to the discarded ($D$) states of the chain, $e$ indicates the environmental subspace, while $n$ stands for the chain index \cite{Anders2005}. Using the NRG representation, the Loschmidt amplitude can be determined from
\begin{equation} \label{Eq:L}
\left< \Psi(t) | \Psi_{0} \right> = \sum_{nse}
\big| \langle \Psi_0|nse\rangle^{\! D} \big|^2 e^{-i E_{ns}^D t},
\end{equation}
where $E_{ns}^D$ denotes the eigenenergy of the state $|nse\rangle^{\!D}$. In calculations we make use of the matrix product state formulation of the problem~\cite{WrzesniewskiWeymann-2019}. Moreover, we assume the discretization parameter satisfies $2 \leqslant \Lambda \leqslant 2.5$, and keep at least $N_K=2000$ states in the computations.

\begin{figure}
\includegraphics[width=0.99\columnwidth]{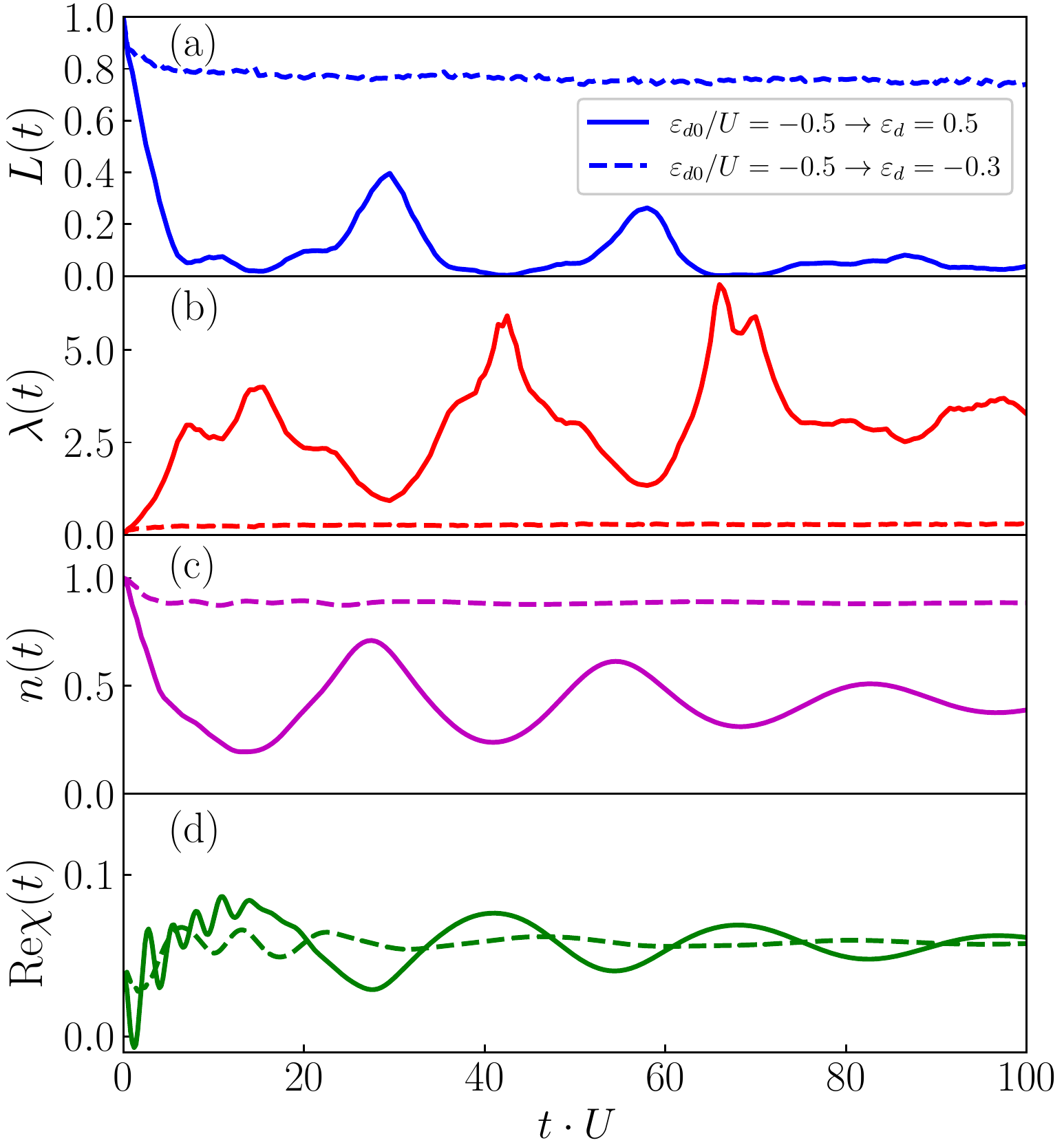}
\caption{(a) The time evolution of the Loschmidt echo,
(b) the return rate,
(c) the dot's occupation, and (d) the on-dot pairing 
after the quench in the position of the quantum dot level, as indicated in the figure. The other parameters are the same as in Fig.~\ref{quench_in_Gamma_S} with $\Gamma_S/U=0.2$.
} 
\label{quench_in_epsilon}
\end{figure}

Let us consider a sudden variation of the coupling $\Gamma_{S}$ for the simple case of a half-filled QD, $\varepsilon_{d}=-U/2$. Figure~\ref{quench_in_Gamma_S} shows the time evolution, when the system is quenched along the dashed line in Fig.~\ref{scheme}(b). The dashed curves correspond to the situation when the quench is performed within the same doublet phase. In that case the Loschmidt echo decays very slowly as the time increases. Consequently, no signatures of DQPTs 
are visible in $\lambda(t)$. On the contrary, when
the quench is performed across the phase boundary 
(see the solid lines in Fig.~\ref{quench_in_Gamma_S}),
$L(t)$ exhibits a pronounced oscillatory behavior
and cusps in the return rate $\lambda(t)$. This is a clear indication of dynamical quantum phase transitions present in the system. Interestingly,
the critical times coincide with appropriate 
oscillations visible in the pairing amplitude shown in Fig.~\ref{quench_in_Gamma_S}(c). The behavior of the return rate around the critical time is presented in Fig~\ref{RR_tc}, where the inset shows the normalized return rate plotted vs dimensionless time. As can be seen, the behavior around the critical time can be described by
\begin{equation}
    \frac{\lambda(t)}{\lambda(t_c)} \sim \frac{|t-t_c|}{t_c}.
\end{equation}

Under stationary conditions one can traverse the
quantum phase transition also by an appropriate variation of the QD energy level. Figure~\ref{quench_in_epsilon} presents the relevant time-dependent quantities obtained for $\Gamma_{S}/U=0.2$ by performing the quench in the dot level $\varepsilon_d$ along the dotted line visible in Fig.~\ref{scheme}(b). Again, the dashed lines correspond to the case when the quench is performed in the doublet phase, to illustrate that the DQPT does not occur then. However, when quenching across the phase boundary clear indications of dynamical quantum critical behavior are present. Note that for this type of quench the cusp-like behavior of $\lambda(t)$
is slightly smeared, which can be attributed to
the fact that now dynamical changes occur
both in the occupation and pairing function.

{\it Summary. --}
We have studied the dynamical properties of the correlated quantum dot sandwiched between the metallic and superconducting leads. Considering the quantum quenches across the the phase boundary between the singlet and doublet phases, triggered by either an abrupt variation of the coupling of the quantum dot to the superconductor or a sudden change of the dot's energy level, we have found clear signatures of dynamical quantum phase transitions. Thus, our work paves the way for the exploration of DQPT in mesoscopic systems, in which the microscopic parameters can be tuned in a fully controllable fashion, allowing for exploration of dynamical critical phenomena with contemporary experimental techniques.

{\it Added note.  --}
At final stage of completing this paper we noticed similar ideas concerning time-dependent transition from a singlet to doublet state for a classical impurity discussed in Ref.~\cite{Morr-2021}.

{\it Acknowledgments.} --
This work is supported by the National Science Centre (Poland) under the grants 2017/27/B/ST3/00621 (KW, IW), 2017/27/B/ST3/01911 (TD), and 2019/35/B/ST3/03625 (NS).

\bibliography{biblio}

\end{document}